\documentclass[11pt]{article}
\usepackage{amsfonts,amsmath}
\def\bbt{\bibitem}
\def\be{\begin{equation}}
\def\en{\end{equation}}
\def\ber{\begin{eqnarray}}
\def\enr{\end{eqnarray}}
\def\nmb{ \nonumber\\}
\def\d{\partial}

\def\ov{\over }
\def\tld{\tilde}

\def\sgm{\sigma}

\def\al{\alpha}
\def\bet{\beta}
\def\gm{\gamma}

\def\im{\imath}

\def\et{\eta}

\def\eps{\epsilon}

\def\ups{\upsilon}

\def\dlt{\delta}

\def\bh{{\bf h}}

\def\bt{{\bf t}}

\def\be{{\bf e}}

\def\bmu{{\boldsymbol \mu}}

\begin{document}
%\nopagenumbers
%\rightline{Landau Tmp.}
\vskip 2 true cm

\centerline{\bf Gepner-like models and Landau-Ginzburg/sigma-model correspondence. }

\vskip 1.5 true cm
\centerline{\bf S. E. Parkhomenko}
\centerline{Landau Institute for Theoretical Physics}
\centerline{142432 Chernogolovka, Russia}
\vskip 0.5 true cm
\centerline{spark@itp.ac.ru}
\vskip 1 true cm
\centerline{\bf Abstract}
\vskip 0.5 true cm

 The Gepner-like models of $k^{K}$-type is considered. When $k+2$ is multiple of
$K$ the elliptic genus and the Euler characteristic is calculated.
Using free-field representation
we relate these models with $\sigma$-models on hypersurfaces in the total space of
anticanonical bundle over the projective space $\mathbb{P}^{K-1}$.

\vskip 10pt

"{\it PACS: 11.25Hf; 11.25 Pm.}"


\smallskip
\vskip 10pt
\centerline{\bf0. Introduction}
\vskip 10pt

 Since the famous work of Gepner \cite{Gep} the geometric aspects underlying his
purely algebraic, Conformal Field Theory (CFT) construction of the superstring vacuum
are the area of intensive studies. His conjecture that there is some relationship between CY 
sigma model and the product of $N=2$ minimal models has been essentially clarified in the works 
\cite{Gep}-\cite{EOTY}, \cite{EllW}-\cite{LSM}. Mirror symmetry, discovered 
in \cite{LVW}, \cite{Di}-\cite{GPl} is one of the most important results of this continuing
line of research.

 In the important work of Borisov \cite{B} the vertex operator algebra endowed with $N=2$
Virasoro superalgebra action has been constructed for each pair of dual reflexive polytopes
defining toric CY manifold. Thus he constructed directly CFT from toric dates of CY manifold.
The approach of Borisov is based essentially on the important 
work of Malikov, Schechtman and Vaintrob \cite{MSV} where a certain sheaf of vertex algebras 
which is called chiral de Rham complex has been introduced. Roughly speaking the construction 
of ~\cite{MSV} is a kind of free-field representation known as $"bc-\bet\gm"$-system which in case of 
$N=2$ superconformal sigma model on toric CY is closely related with 
the Feigin and Semikhatov free-field representation ~\cite{FeS} of $N=2$ supersymmetric 
minimal models.  This circumstance is probably the key to understanding string geometry of 
Gepner models and proving Gepner's conjecture.

 The significant step in this direction has been made in the paper~\cite{GobM} where
the vertex algebra of certain Landau-Ginzburg (LG) orbifold has been related to chiral de Rham
complex of toric CY manifold by some spectral sequence. The CY manifold has been realized
as an algebraic surface degree $K$ in the projective space $\mathbb{P}^{K-1}$ and one of the key points 
of ~\cite{GobM} is that the free-field representation of the corresponding LG orbifold is given 
by $K$ copies of $N=2$ minimal model free-field representation of \cite{FeS}.

 In this note we consider Gepner-like models which are the products of $N=2$ minimal
models projected by the integer $U(1)$ charge condition. Thus we orbifoldize
the product of $N=2$ minimal models in complete similarity to the case of Gepner models.
The only difference is that we relax the total central charge condition for the product
of minimal models and consider the product of $K$-copies of $N=2$ minimal models with
equal central charges $c_{1}=...=c_{K}={3k\ov k+2}$, where $k+2$ is multiple of $K$.
When $k+2=K$ we are in the CY situation considered in \cite{GobM}. In general case we calculate 
in Sect.1. the elliptic genus and Euler characteristic of the model. 
In Sect.2. we use free-field representation
of \cite{FeS} to relate this model with $\mathbb{C}^{K}/\mathbb{Z}_{k+2}$ LG orbifold. 
In Sect.3. we discuss briefly the resolution of orbifold singularity and relate the model with
$\sigma$-model on a hypersurface in the total space of anticanonical bundle over the
projective space $\mathbb{P}^{K-1}$.

\vskip 10pt
\centerline{\bf 1. The Elliptic genus and Euler characteristic of the Gepner-like models.}
\vskip 10pt

 In this section the elliptic genus is calculated for certain orbifold of the product of $N=2$
minimal models. As a preliminary
we represent a collection of known facts on the $N=2$ minimal models and fix the notations.

\leftline {\bf 1.1. The products of $N=2$ minimal models.}

 The tensor product of $K$ $N=2$ unitary minimal models can be characterized
by $K$ dimensional vector $\bmu=(\mu_{1},...,\mu_{K})$,
where $\mu_{i}\geq 2$ being integer defines the central charge of the individual model
by $c_{i}=3(1-{2\ov\mu_{i}})$. 
For each individual minimal model we denote by $M_{h,t}$ the irreducible unitary 
$N=2$ Virasoro superalgebra representation in NS sector and denote by 
$\chi_{h,-t}(q,u)$ the character of the representation, where
$h=0,...,\mu-2$ and $t=0,...,h$. 
There are the following important automorphisms of the irreducible modules and characters
~\cite{FeS},~\cite{FeSST}.
\ber
M_{h,t}\equiv M_{\mu-h-2,t-h-1}, \
\chi_{h,t}(q,u)=\chi_{\mu-h-2,t-h-1}(q,u),
\label{1.reflect}
\enr
\ber
M_{h,t}\equiv M_{h,t+\mu}, \ \chi_{h,t+\mu}(q,u)=
\chi_{h,t}(q,u),
\label{1.oddper}
\enr
where $\mu$ is odd and
\ber
M_{h,t}\equiv M_{h,t+\mu}, \ \chi_{h,t+\mu}(q,u)=
\chi_{h,t}(q,u),\ h\neq [{\mu\ov2}]-1, \nmb
M_{h,t}\equiv M_{h,t+[{\mu\ov2}]}, \
\chi_{h,t+[{\mu\ov 2}]}(q,u)=\chi_{h,t}(q,u),\ h=[{\mu\ov2}]-1,
\label{1.evper}
\enr
where $\mu$ is even.
In what follows we extend the set of admissible $t$: 
\ber
t=0,...,\mu-1
\label{1.textend}
\enr
using the automorphisms above.

 The parameter $t\in Z$ labels the spectral flow automorphisms \cite{SS} of $N=2$
Virasoro superalgebra in NS sector
\ber
G^{\pm}[r]\rightarrow G_{t}^{\pm}[r]\equiv U^{t}G^{\pm}[r]U^{-t}\equiv G^{\pm}[r\pm t], \nmb
L[n]\rightarrow L_{t}[n]\equiv U^{t}L[n]U^{-t}\equiv L[n]+t J[n]+t^{2}{c\ov 6}\dlt_{n,0},
\nmb
J[n]\rightarrow J_{t}[n]\equiv U^{t}J[n]U^{-t}\equiv J[n]+t {c\ov 3}\dlt_{n,0},
\label{1.flow}
\enr
where $U^{t}$ denotes the spectral flow operator generating twisted sectors and $r$ is half-integer for the modes of
the spin-$3/2$ fermionic currents $G^{\pm}(z)$ while $n$ is integer for the modes of stress-energy 
tensor $T(z)$ and $U(1)$-current $J(z)$ of the $N=2$ Virasoro superalgebra. 
So allowing $t$ to be half-integer we recover the irreducible
representations and characters in the $R$ sector.

 We use the following expression for the characters found in \cite{FeSST}
\ber
\chi_{h,-t}(u,q)=q^{{h\ov 2\mu}+{c\ov 6}t^{2}+{th\ov\mu}-{c\ov 24}}q^{{1-\mu\ov 8}}
u^{{h\ov\mu}+{ct\ov 3}}
({\et(q^{\mu})\ov\et(q)})^{3}
\nmb
\prod_{n=0}{(1+uq^{{1\ov 2}+t+n})\ov (1+u^{-1}q^{-{1\ov 2}-t+n\mu})}
{(1+u^{-1}q^{{1\ov 2}-t+n})\ov (1+uq^{{1\ov 2}+t+(n+1)\mu})}
{(1-q^{n+1})\ov (1-q^{(n+1)\mu})}
\nmb
\prod_{n=0}{(1-q^{-1-h+n\mu})\ov (1+uq^{-{1\ov 2}-h+t+n\mu})}
{(1-q^{1+h+(n+1)\mu})\ov (1+u^{-1}q^{{1\ov 2}+h-t+(n+1)\mu})}
\label{1.char}
\enr
where
\ber
\et(q)=q^{{1\ov 24}}\prod_{n=1}(1-q^{n})
\label{1.Ded}
\enr

 The $N=2$ Virasoro superalgebra generators in the product of minimal models are given by the sums of
generators of each minimal model
\ber
G^{\pm}[r]=\sum_{i}G^{\pm}_{i}[r], \nmb
J[n]=\sum_{i}J_{i}[n], \
T[n]=\sum_{i}T_{i}[n],
\nmb
c=\sum_{i}3(1-{2\ov\mu_{i}})
\label{1.Vird}
\enr
This algebra is obviously acting in the tensor products 
$M_{\bh,\bt}=\otimes_{i=1}^{K}M_{h_{i},t_{i}}$ of the irreducible $N=2$ Virasoro superalgebra 
representations of each individual model. We use the similar notation for the corresponding 
product of characters
\ber
\chi_{\bh,\bt}(q,u)=\prod_{i=1}^{K}\chi_{h_{i},t_{i}}(q,u)
\label{1.prodchi}
\enr

 By the definition ~\cite{EllW} the elliptic genus of $N=2$ supersymmetric CFT is given by
\ber
Ell(\tau,\ups )=
\nmb
Tr_{(R\times R)}((-1)^{f+\bar{f}}
\exp{[\im 2\pi\tau (L[0]-{c\ov 24})+\im 2\pi\ups (J[0]-{c\ov 6})]}
\exp{[\im 2\pi\bar{\tau} (\bar{L}[0]-{c\ov 24})]})
\label{1.Ellg}
\enr
The trace is taken over the Hilbert space in $R\times R$ sector and the
operators $f$ and $\bar{f}$ are fermion number operators in left-moving and right-moving
sectors.

\leftline {\bf 1.2. Elliptic genus calculation.}

 Now we calculate the elliptic genus for the case of orbifold of the product
of minimal models when $K$-dimensional vector 
is given by $\bmu=(\mu,...,\mu)$, where $\mu$ is positive and multiple of $K$.
In these models the total central charge is $3K(1-{2\ov\mu})$, so it is no longer integer 
and multiple of 3 in general.
Except the cases $\mu=K,2K$ they can not be considered in general as the models of superstring 
compactification.
Nevertheless the orbifold projection consistent with modular invariance still exists \cite{V} 
which makes them to be interesting $N=2$ supesymmetric models of CFT from geometric point of view.
The general prescription for the orbifold elliptic genus calculation has been developed 
in \cite{KYY} which we shall follow closely.

 Before the orbifold projection the elliptic genus of the product of $N=2$ minimal models
can be calculated as the elliptic genus of the LG-model \cite{EllW}, \cite{KYY}.
\ber
Ell(\tau, \ups)=\prod_{i=1}^{K}Ell_{i}(\tau,\ups),
\nmb
Ell_{i}(\tau,\ups)=u^{-{c_{i}\ov 6}}{(1-u^{1-{1\ov\mu}})\ov (1-u^{1\ov\mu})}
\prod_{n=1}{(1-u^{1-{1\ov\mu}}q^{n})\ov (1-u^{{1\ov\mu}}q^{n})}
{(1-u^{-1+{1\ov\mu}}q^{n})\ov (1-u^{-{1\ov\mu}}q^{n})}
\label{1.ElLG}
\enr
In fact one can get this expression directly using free-field realization of $N=2$ minimal model
of Section 2 giving thereby the proof of LG-calculation from \cite{EllW}.

 The orbifold group is $\mathbb{Z}_{\mu}$ and generated by 
\ber
g=\exp (\im 2\pi J[0])
\label{1.gorb}
\enr
According to \cite{KYY} the orbifold elliptic genus is given by
\ber
Ell_{orb}(\tau,\ups )=
\nmb
{1\ov \mu}\sum_{n,l=0}^{\mu-1}\eps(n,l)\exp{(\im 2\pi {c\ov 6}nl)} 
\prod_{i=1}^{K}\exp{(\im 2\pi{c_{i}\ov 6}(n^{2}\tau+2n\ups ))}Ell_{i}(\tau,\ups+n\tau+l )
\label{1.Ellorb}
\enr
where
\ber
\eps(n,l)=\exp{(\im \pi (n+l+nl)K)}
\label{1.cocycl}
\enr

The summation over $n$ is due to the spectral flow twisted sector
generated by the product of spectral flow twisted operators $\prod_{i=1}^{K}U_{i}^{n}$.
The summation over $l$ corresponds to the projection on the $\mathbb{Z}_{\mu}$-invariant states. 
The Ramound
sector is given by the ${1\ov 2}$-twisted sector. By this convention the chiral-primary fields
of NS sector corresponds to the ground states in R sector. 

 The Euler characteristic is given by the value of the elliptic genus at $\ups=0$. 
\ber
Eu\equiv \lim_{\ups\rightarrow 0}Ell_{orb}(\tau, \ups)=
{(\mu-1)^{K}\ov\mu}+(-1)^{K}{\mu^{2}-1\ov\mu}
\label{1.Eu}
\enr
This expression follows from 
\ber
\lim_{\ups\rightarrow 0}Ell_{i}(\tau,\ups+n\tau+l)=(-1)^{n}(\mu_{i}-1)
\exp{(-\im 2\pi{c_{i}\ov 6}n^{2})},\ if \ l=0,
\nmb
\lim_{\ups\rightarrow 0}Ell_{i}(\tau,\ups+n\tau+l)=(-1)^{n+l+1}
\exp{(\im 2\pi{nl\ov\mu_{i}})}\exp{(-\im 2\pi{c_{i}\ov 6}n^{2})}, \ if \ l>0.
\label{1.Eunm}
\enr

\vskip 10pt
\centerline {\bf 2. LG orbifold geometry of Gepner-like models.}
\vskip 10pt

 In this section we relate the Gepner-like models to the LG orbifolds 
$\mathbb{C}^{K}/\mathbb{Z}_{\mu}$ using essentially the free-field construction of 
irreducible representations of $N=2$ minimal models
found by Feigin and Semikhatov in \cite{FeS}. 

\leftline{\bf 2.1. Free-field realization of $N=2$ minimal models.}

 Let $X(z), X^{*}(z)$ be the free bosonic fields and $\psi(z), \psi^{*}(z)$ be the
free fermionic fields (in the left-moving sector) 
so that its OPE's are given by
\ber
X^{*}(z_{1})X(z_{2})=\ln(z_{12})+reg.,\nmb
\psi^{*}(z_{1})\psi(z_{2})=z_{12}^{-1}+reg,
\label{2.ope}
\enr
where $z_{12}=z_{1}-z_{2}$. Then for an arbitrary number
$\mu$ the currents of $N=2$ super-Virasoro
algebra are given by
\ber
G^{+}(z)=\psi^{*}(z)\d X(z) -{1\ov \mu} \d \psi^{*}(z), \
G^{-}(z)=\psi(z) \d X^{*}(z)-\d \psi(z), \nmb
J(z)=\psi^{*}(z)\psi(z)+{1\ov \mu}\d X^{*}(z)-\d X(z), \nmb
T(z)=\d X(z)\d X^{*}(z)+
{1\ov 2}(\d \psi^{*}(z)\psi(z)-\psi^{*}(z)\d \psi(z))-\nmb
{1\ov 2}(\d^{2} X(z)+{1\ov \mu}\d^{2} X^{*}(z)),
\label{2.min}
\enr
and the central charge is
\ber
c=3(1-{2\ov \mu}).
\label{2.cent}
\enr

  As usual, the fermions are expanded into the
half-integer modes in NS sector and they are expanded into integer modes in R sector
\ber
\psi(z)=\sum_{r}\psi[r]z^{-{1\ov 2}-r},\
\psi^{*}(z)=\sum_{r}\psi^{*}[r]z^{-{1\ov 2}-r},\
G^{\pm}(z)=\sum_{r}G^{\pm}[r]z^{-{3\ov 2}-r},
\label{2.NSR}
\enr
The bosons are expanded in both sectors into the integer
modes:
\ber
\d X(z)=\sum_{n\in Z}X[n]z^{-1-n},\
\d X^{*}(z)=\sum_{n\in Z}X^{*}[n]z^{-1-n},\nmb
J(z)=\sum_{n\in Z}J[n]z^{-1-n},\
T(z)=\sum_{n\in Z}L[n]z^{-2-n}.
\label{2.NSRb}
\enr

 In NS sector $N=2$ Virasoro superalgebra is acting naturally in Fock module
$F_{p,p^{*}}$ generated by the fermionic
operators $\psi^{*}[r]$, $\psi[r]$, $r<{1\ov2}$, and bosonic operators
$X^{*}[n]$, $X[n]$, $n<0$ from the vacuum state $|p,p^{*}>$ such that
\ber
\psi[r]|p,p^{*}>=\psi^{*}[r]|p,p^{*}>=0, r\geq {1\ov 2},\nmb
X[n]|p,p^{*}>=X^{*}[n]|p,p^{*}>=0, n\geq 1, \nmb
X[0]|p,p^{*}>=p|p,p^{*}>, \
X^{*}[0]|p,p^{*}>=p^{*}|p,p^{*}>.
\label{2.vac}
\enr
It is a primary state with respect
to the $N=2$ Virasoro algebra
\ber
G^{\pm}[r]|p,p^{*}>=0, r>0, \nmb
J[n]|p,p^{*}>=L[n]|p,p^{*}>=0, n>0, \nmb
J[0]|p,p^{*}>={j\ov \mu}|p,p^{*}>=0, \nmb
L[0]|p,p^{*}>={h(h+2)-j^{2}\ov 4\mu}|p,p^{*}>=0,
\label{2.hwv}
\enr
where $j=p^{*}-\mu p$, $h=p^{*}+\mu p$.

 When $\mu-2$ is integer and non negative the Fock modules are highly reducible representations 
of $N=2$ Virasoro algebra.

The irreducible module $M_{h,j}$ is given by cohomology of some complex building up from 
Fock modules. This complex has been constructed in ~\cite{FeS}. Let us consider first free-field 
construction for the chiral module $M_{h,0}$. In this case the complex
(which is known due to Feigin and Semikhatov as butterfly resolution)
can be represented by the following diagram
\ber
\begin{array}{ccccccccccc}
&&\vdots &\vdots &&&&&&\\
&&\uparrow &\uparrow &&&&&&\\
\ldots &\leftarrow &F_{1,h+\mu} &\leftarrow
F_{0,h+\mu}&&&&&&\\
&&\uparrow &\uparrow &&&&&&\\
\ldots &\leftarrow &F_{1,h} &\leftarrow F_{0,h}&&&&&&\\
&&&&\nwarrow&&&&&\\
&&&&&F_{-1,h-\mu}&\leftarrow &F_{-2,h-\mu}&\leftarrow&\ldots\\
&&&&&\uparrow &&\uparrow&\\
&&&&&F_{-1,h-2\mu}&\leftarrow &F_{-2,h-2\mu}&\leftarrow&\ldots\\
&&&&&\uparrow &&\uparrow &&\\
&&&&&\vdots &&\vdots &&
\end{array} \nmb
\label{2.but}
\enr
The horizontal arrows in this diagram are given by the action of
\ber
Q^{+}=\oint dz S^{+}(z), \ S^{+}(z)=\psi^{*}\exp(X^{*})(z),
\label{2.chrg+}
\enr
The vertical arrows are given by the action of 
\ber
Q^{-}=\oint dz S^{-}(z), \ S^{-}(z)=\psi\exp(\mu X)(z),
\label{2.chrg-}
\enr
The diagonal arrow at the middle of butterfly resolution
is given by the action of $Q^{+}Q^{-}$. It is a complex due to the following properties
screening charges $Q^{\pm}$
\ber
(Q^{+})^{2}=(Q^{-})^{2}=\{Q^{+},Q^{-}\}=0.
\label{2.BRST}
\enr

 The main statement of ~\cite{FeS} is that the complex (\ref{2.but}) is exact
except at the $F_{0,h}$ module, where the cohomology is given by
the chiral module $M_{h,0}$.

 To get the resolution for the irreducible module $M_{h,t}$
one can use the observation ~\cite{FeS} that all irreducible modules can be obtained
from the chiral module $M_{h,0}$, $h=0,...,\mu-2$ by the spectral flow
action $U^{-t}, t=1,...,\mu-1$. The spectral flow action on the free
fields can be easily described if we bosonize fermions $\psi^{*}, \psi$
\ber
\psi(z)=\exp(-\phi(z)), \ \psi^{*}(z)=\exp(\phi(z)).
\label{2.fbos}
\enr
and introduce spectral flow vertex operator
\ber
U^{t}(z)=\exp(-t(\phi+{1\ov \mu}X^{*}-X)(z)).
\label{2.vflow}
\enr

Using the resolution (\ref{2.but}) one can get directly the expression (\ref{1.ElLG}) for the elliptic genus. 
By the spectral flow we obtain also the 
expression (\ref{1.char}) for the character.
 
 The resolutions and irreducible modules in R sector are
generated from the resolutions and modules in NS sector by the
spectral flow operator $U^{{1\ov2}}$.

\leftline {\bf 2.2.Free-field realization of the product of
minimal models.}

 It is clear how to generalize the free-field representation for the case of
tensor product of $K$ $N=2$ minimal models.
One has to introduce (in the left-moving sector)
the free bosonic fields $X_{i}(z), X^{*}_{i}(z)$ and free
fermionic fields $\psi_{i}(z), \psi^{*}_{i}(z)$, $i=1,...,K$
so that its singular OPE's are given by (\ref{2.ope}).
The $N=2$ superalgebra Virasoro currents for each of the models are given by (\ref{2.min}).
To describe the products of irreducible representations $M_{\bh,\bt}$ we introduce
the fermionic screening currents and their charges
\ber
S^{+}_{i}(z)=\psi^{*}_{i}\exp(X^{*}_{i})(z), \nmb
S^{-}_{i}(z)=\psi_{i}\exp(\mu_{i}X_{i})(z), \nmb
Q^{\pm}_{i}=\oint dz S^{\pm}_{i}(z).
\label{2.chrgi}
\enr
Then the module $M_{\bh,0}$ is given by the cohomology of the product
of butterfly resolutions (\ref{2.but}) for each minimal model. The resolution
of the module $M_{\bh,\bt}$ is generated by the spectral flow operator
$U^{\bt}=\prod_{i}U_{i}^{t_{i}}$, $t_{i}=1,...,\mu_{i}-1$, where
$U_{i}^{t_{i}}$ is the spectral flow operator from the $i$-th minimal model (\ref{2.vflow}). 
Allowing $t_{i}$ to be half-integer we generate the corresponding objects in R sector.

\leftline {\bf 2.3. LG orbifold geometry of Gepner-like models.}

 The elliptic genus (\ref{1.Ellorb}) can be considered as the Euler character
of certain complex. It is an orbifold of the complex which is given by the sum 
of products of butterfly resolutions for the modules $M_{\bh,0}$. 
The cohomology of this complex can be calculated by two steps.

 At first step we take the cohomology with respect to the operator 
\ber
Q^{+}=\sum_{i=1}^{K}Q^{+}_{i}
\label{2.Qplus}
\enr
It is generated by $bc\bet\gm$ system of fields
\ber
a_{i}(z)=\exp{[X_{i}]}(z),\ \al_{i}(z)=\psi_{i}\exp{[X_{i}]}(z),
\nmb
a^{*}_{i}(z)=(\d X_{i}^{*}-\psi_{i}\psi^{*}_{i})
\exp{[-X_{i}]}(z), \
\al^{*}_{i}(z)=\psi^{*}_{i}\exp{[-X_{i}]}(z)
\label{2.btgm}
\enr
The fields $a_{i}(z)$ correspond to the coordinates $a_{i}$ on the complex space
$\mathbb{C}^{K}$, the fields $a^{*}_{i}(z)$ correspond to the operators ${\d \ov \d a_{i}}$.
The fields $\al_{i}(z)$ correspond to the differentials $da_{i}$, while $\al^{*}_{i}(z)$
correspond to the conjugated to $da_{i}$. 

In terms of the fields (\ref{2.btgm}) the N=2 Virasoro superalgebra currents (\ref{1.Vird})
are given by
\ber
G^{-}=\sum_{i}\al_{i}a^{*}_{i}, \
G^{+}=\sum_{i}(1-{1\ov\mu})\al^{*}_{i}\d
a_{i}-{1\ov\mu}a_{i}\d\al^{*}_{i}, \nmb
J=\sum_{i}(1-{1\ov\mu})\al^{*}_{i}\al_{i}+{1\ov\mu}a_{i}a^{*}_{i},\nmb
T=\sum_{i}{1\ov2}((1+{1\ov\mu})\d\al^{*}_{i}\al_{i}-
(1-{1\ov\mu})\al^{*}_{i}\d\al_{i})+
(1-{1\ov2\mu})\d a_{i}a^{*}_{i}-
{1\ov2\mu}a_{i}\d a^{*}_{i}
\label{2.btgmvir}
\enr
Notice that zero mode $G^{-}[0]$ is acting on the space of states generated by $bc\bet\gm$ system
of fields similar to the de Rham differential action on the de Rham complex of $\mathbb{C}^{K}$.
The next important property is the behaviour of the $bc\bet\gm$ system under the the change of
coordinates on \cite{MSV}. It endows the $bc\bet\gm$ system (\ref{2.btgm})
with the structure of sheaf known as chiral de Rham complex due to \cite{MSV}. It provides the geometric 
meaning to the algebraic $\mathbb{Z}_{\mu}$-orbifod projection of the product of minimal models. 

 Indeed, the screening charges $Q^{+}_{i}$ correspond
to some cone in the lattice $\mathbb{Z}^{K}$ generated by the basic vectors $e_{i}$.  
The monomials generated by the fields $a_{i}(z)$ correspond to
the dual cone in the dual lattice \cite{Ful}. The charges of the fields (\ref{2.btgm}) are given by
\ber
J(z_{1})a_{i}(z_{2})=z_{12}^{-1}{1\ov\mu}a_{i}(z_{2})+r.,
\
J(z_{1})a^{*}_{i}(z_{2})=-z_{12}^{-1}{1\ov\mu}a^{*}_{i}(z_{2})+r.,
\nmb
J(z_{1})\al_{i}(z_{2})=-z_{12}^{-1}(1-{1\ov\mu})\al_{i}(z_{2})+r.,\
J(z_{1})\al^{*}_{i}(z_{2})=z_{12}^{-1}(1-{1\ov\mu})\al^{*}_{i}(z_{2})+r.
\label{2.jchrg}
\enr

 Hence, making the projection on $\mathbb{Z}_{\mu}$-invariant states and adding twisted sectors 
generated by $\prod_{i=1}^{\mu-1}(U_{i})^{n}$ we obtain toric construction of the chiral 
de Rham complex of the orbifold $\mathbb{C}^{K}/\mathbb{Z}_{\mu}$.
The chiral de Rham complex on the orbifold has recently been introduced in \cite{FrSc}.

 The second step in the cohomology calculation is given by
the cohomology with respect to the differential
$Q^{-}=\sum_{i=1}^{K}Q^{-}_{i}$.
This operator survives the orbifold projection and its expression in terms of fields
(\ref{2.btgm}) is
\ber
Q^{-}=\oint dz \sum_{i=1}^{K}\al_{i}(a_{i})^{\mu-1}
\label{2.Qminus1}
\enr
Therefore the second step of cohomology calculation gives the restriction of the space of states 
to the points $dW=0$ of the potential
\ber
W=\sum_{i=1}^{K}(a_{i})^{\mu}
\label{2.LG}
\enr
Thus the total space of states is the space of states of LG orbifold 
$\mathbb{C}^{K}/\mathbb{Z}_{\mu}$
and the expression (\ref{1.Ellorb}) is the elliptic genus of this LG orbifold. 

\vskip 10pt
\centerline {\bf 3. LG/sigma-model correspondence conjecture.}
\vskip 10pt

 As it has already been mentioned the case of $\mu=K$ corresponds to CY manifold
which is given by degree $K$ surface in projective space $\mathbb{P}^{K-1}$. The 
chiral de Rham complex on this manifold has been constructed in \cite{B},~\cite{GobM}.
In \cite{GobM} the chiral de Rham complex on the CY manifold in $\mathbb{P}^{K-1}$ has 
been calculated by the spectral sequence which relates this complex to the chiral 
de Rham complex on the LG orbifold.

 We briefly consider here the spectral sequence of \cite{GobM} for the simplest case of 
0-dimensional CY manifold in $\mathbb{P}^{1}$ which corresponds to $\bmu=(2,2)$ model. Then we
consider the possible generalization to the case when $\mu$ is multiple
of $K$ and discuss the underlying geometry.

 When $K=2$ and $\bmu=(2,2)$ the expression $(\ref{1.Ellorb})$ gives the elliptic genus
of the LG orbifold $\mathbb{C}^{2}/\mathbb{Z}_{2}$ with the potential 
\ber
W=a_{1}^{2}+a_{2}^{2}
\label{3.lgpot}
\enr 
as we have seen in Sect.2.

According to the construction  \cite{B},~\cite{GobM} the resolution of 
the orbifold singularity is given by the screening charge
\ber
D^{+}_{0}=\oint dz {1\ov 2}(\psi^{*}_{1}+\psi^{*}_{2})\exp({1\ov 2}(X^{*}_{1}+X^{*}_{2}))(z)
\label{3.Q0}
\enr
It gives a fan \cite{Ful} consisting of two 2-dimensional cones $\sgm_{1}$ and $\sgm_{2}$, 
generated in the lattice
$({1\ov 2}\mathbb{Z})^{2}$ by the vectors $(e_{1},{1\ov 2}(e_{1}+e_{2}))$ and vectors
$(e_{2},{1\ov 2}(e_{1}+e_{2}))$ correspondingly. To each of the cones $\sgm_{i}$
the $bc\bet\gm$ system of fields is related by the cohomology of 
the differential $Q^{+}_{i}+Q^{+}_{0}$ (the first step of cohomology calculation). 
By the explicit calculations (see for example \cite{B})
one can show that these two systems generate the space of sections of the chiral de Rham
complex on the open sets of the standard covering of the total space of $O(2)$ line bundle 
over $\mathbb{P}^{1}$.  The Chech complex of the standard
covering glues these sections into the chiral de Rham complex of the total space of the bundle.
The cohomology with respect to the differential $Q^{-}$ restricts the complex to the set of points
$dW=0$.

 Now we propose the orbifold singularity resolution when
$K=2$ and $\mu=2m$, $m=1,2,...$. In this case we have LG orbifold 
$\mathbb{C}^{2}/\mathbb{Z}_{2m}$
with the potential
\ber
W=a_{1}^{2m}+a_{2}^{2m}
\label{3.lgpot1}
\enr
To resolve the orbifold singularity we consider the following set of screening charges
\ber
D^{+}_{n}=\oint dz ({m-n\ov 2m}\psi^{*}_{1}+{m+n\ov 2m}\psi^{*}_{2})
\exp({m-n\ov 2m}X^{*}_{1}+{m+n\ov 2m}X^{*}_{2})(z),\ n=-m+1,...,m-1
\label{3.Qn}
\enr 
It is easy to check that these operators commute with the total $N=2$ Virasoro superalgebra
currents (\ref{2.btgmvir}). They commute also with the operators $Q^{-}_{i}$ when $\mu=2m$.
But most of the fields (\ref{3.Qn}) can not appear as marginal 
operators of the model because they should come from twisted sectors which are not exist 
in the model.
The only exception comes from the spectral flow operator $\prod_{i=1}^{\mu-1}(U_{i})^{n}$. Hence 
the only screening charge one can add to resolve the singularity is $D^{+}_{0}$, the middle one 
from (\ref{3.Qn}). By this means we are turning back to the fan of $\bmu=(2,2)$ model. 
The important difference however is that the group $\mathbb{Z}_{m}$ is acting on 
the chiral de Rham complex sections. 
But the only $bc\bet\gm$ fields charged with respect to this group correspond to the fibers 
of the $O(2)$-bundle. In other words, the group $\mathbb{Z}_{m}$ is acting only along 
the fibers, so that the base $\mathbb{P}^{1}$ is the fixed point set 
of the action. Therefore we obtain after the blow-up
the $\mathbb{Z}_{m}$-orbifold of the chiral de Rham complex of the $O(2)$-bundle total space.
 
 The differential $Q^{-}$ of the second step cohomology calculation commutes with $D^{+}_{0}$
and survives $\mathbb{Z}_{m}$-projection. It defines the function (potential) $W$ on the 
total space of $O(2)$-bundle and $Q^{-}$-cohomology calculation restricts the
chiral de Rham complex to the $dW=0$ point set of the function.
 
 We find the potential by the explicit calculation in some coordinates. According to the construction
\cite{B} the set of 
sections of chiral de Rham
complex of $O(2)$-bundle over the one of the open set $\Gamma _{i}$ ($i=1,2$) of the 
standard covering of the total space of the bundle
is given by the cohomology of $Q^{+}_{i}+D^{+}_{0}$. For example the sections of chiral de Rham 
complex over the $\Gamma_{1}$ is given by $Q^{+}_{1}+D^{+}_{0}$ 
cohomology and generated by the following $bc\bet\gm$ fields
\ber
b_{0}(z)=\exp{[2X_{2}]}(z),\ \bet_{0}(z)=2\psi_{2}\exp{[2X_{2}]}(z),
\nmb
b^{*}_{0}(z)=({1\ov 2}(\d X_{1}^{*}+\d X_{2}^{*})-2\psi_{2}{1\ov 2}(\psi^{*}_{1}+\psi^{*}_{2}))
\exp{[-2X_{2}]}(z), 
\nmb
\bet^{*}_{0}(z)={1\ov 2}(\psi^{*}_{1}+\psi^{*}_{2})\exp{[-2X_{2}]}(z),
\nmb
b_{1}(z)=\exp{[X_{1}-X_{2}]}(z),\ \bet_{1}(z)=(\psi_{1}-\psi_{2})\exp{[X_{1}-X_{2}]}(z),
\nmb
b^{*}_{1}(z)=(\d X_{1}^{*}-(\psi_{1}-\psi_{2})\psi^{*}_{1})
\exp{[-X_{1}+X_{2}]}(z), \
\bet^{*}_{1}(z)=\psi^{*}_{1}\exp{[-X_{1}+X_{2}]}(z)
\label{3.btgm01}
\enr
Then the potential (\ref{3.lgpot}) takes the form
\ber
W=(b_{0})^{m}(1+(b_{1})^{2m})
\label{3.W01}
\enr
The $dW=0$ points are given by the equations
\ber
(b_{0})^{m-1}=0, \ when \ b_{1}^{2m}\neq -1,
\nmb
(b_{0})^{m}=0, \ when \ b_{1}^{2m}=-1
\label{3.dW01}
\enr

 Analogously, the sections of chiral de Rham 
complex over the $\Gamma_{2}$ are given by the $Q^{+}_{2}+D^{+}_{0}$ cohomology and
generated by the fields
\ber
\tld{b}_{0}(z)=\exp{[2X_{1}]}(z),\ \tld{\bet}_{0}(z)=2\psi_{1}\exp{[2X_{1}]}(z),
\nmb
\tld{b}^{*}_{0}(z)=({1\ov 2}(\d X_{1}^{*}+\d X_{2}^{*})-2\psi_{1}{1\ov 2}(\psi^{*}_{1}+\psi^{*}_{2}))
\exp{[-2X_{1}]}(z), 
\nmb
\tld{\bet}^{*}_{0}(z)={1\ov 2}(\psi^{*}_{1}+\psi^{*}_{2})\exp{[-2X_{1}]}(z),
\nmb
\tld{b}_{1}(z)=\exp{[-X_{1}+X_{2}]}(z),\ \tld{\bet}_{1}(z)=-(\psi_{1}-\psi_{2})\exp{[-X_{1}+X_{2}]}(z),
\nmb
\tld{b}^{*}_{1}(z)=(\d X_{2}^{*}-(-\psi_{1}+\psi_{2})\psi^{*}_{2})
\exp{[X_{1}-X_{2}]}(z), \
\tld{\bet}^{*}_{1}(z)=\psi^{*}_{2}\exp{[X_{1}-X_{2}]}(z)
\label{3.btgm02}
\enr
In these coordinates the potential takes the form
\ber
W=(\tld{b}_{0})^{m}(1+(\tld{b}_{1})^{2m})
\label{3.W02}
\enr
so that $dW=0$ points set is given similar to (\ref{3.dW01}).

 Comparing the expressions (\ref{3.btgm01}) and (\ref{3.btgm02}) we see that
field $b_{0}(z)$ ($\tld{b}_{0}(z)$), corresponds to the coordinate along the fiber and the field 
$b_{1}(z)$ ($\tld{b}_{1}(z)$) corresponds to the 
coordinate along the base $\mathbb{P}^{1}$ of $O(2)$-bundle in the open set $\Gamma_{1}$ ($\Gamma_{2}$).

 For general values of $K$
and $\mu=mK$ the situation is similar. The only screening charge one can add to resolve 
the orbifold singularity comes from the spectral flow operator
\ber
D^{+}_{0}=\oint dz {1\ov K}(\sum_{i}\psi^{*}_{i})
\exp({1\ov K}\sum_{i}X^{*}_{i})(z)
\label{3.QK}
\enr 
Together with $Q^{+}_{i}$ it gives the standard fan of the $O(K)$-bundle total space over $\mathbb{P}^{K-1}$.
The highest dimensional cones $\sgm_{i}$ of the fan are labeled by the sets 
$(D^{+}_{0},Q^{+}_{1},...,Q^{+}_{i-1},Q^{+}_{i+1},...,Q^{+}_{K})$, where $Q^{+}_{i}$ is missing.
The group 
$\mathbb{Z}_{m}$ is acting along the fibers of the bundle with the fixed point set $\mathbb{P}^{K-1}$.
Thus we obtain after the blow-up
the $\mathbb{Z}_{m}$-orbifold of the chiral de Rham complex of the $O(K)$-bundle total space.
The differential $Q^{-}$ commutes with $D^{+}_{0}$ due to the condition $\mu=Km$ and
survives $\mathbb{Z}_{m}$-projection hence, it defines the potential (\ref{2.LG}) on the total
space of the $O(K)$-bundle. Therefore the expression (\ref{1.Ellorb}) is the elliptic genus
of the chiral de Rham complex
of the $O(K)$-bundle restricted to the set of points $dW=0$:
\ber
(b_{0})^{m-1}=0, \ when \ \sum_{i=1}^{K-1}b_{i}^{Km}\neq -1,
\nmb
(b_{0})^{m}=0, \ when \ \sum_{i=1}^{K-1}b_{i}^{Km}=-1
\label{3.dWK}
\enr
where $b_{0}$ is the coordinate along the fiber and $b_{i}$, $i=1,...K-1$ are the coordinates along the
base $\mathbb{P}^{K-1}$ in some of the open set of the standard covering of the $O(K)$-bundle.
The algebraic manifold determined by the equations (\ref{3.dWK}) is singular except the case $m=1$.
Nevertheless the Euler characteristics (\ref{1.Eu}) can be represented in the form compatible
with these equations:
\ber
Eu=(-1)^{K}(K+{(1-mK)^{K}-1\ov mK}+(m-1)K)=
\nmb
(-1)^{K}(mEu(V)+(m-1)Eu(\mathbb{P}^{K-1}\setminus  V))
\label{3.EuK}
\enr
where $V$ is the set of points in $\mathbb{P}^{K-1}$ satisfying the equation
$\sum_{i=1}^{K-1}b_{i}^{Km}+1=0$.

 More detailed investigation of toric geometry of the models we left for the future.

\vskip 20pt
\leftline{\bf Acknowledgements}
\vskip 10pt

I thank Boris Feigin and Mikhail Bershtein for helpful discussions.
I especially thank Fyodor Malikov for collaboration on the results in Section 3.
This work was supported in part by grants RFBR-07-02-00799-a, SS3472.2008.2,
RFBR-CNRS (PICS) 09-02-91064.

\end{document}